\documentclass[12pt, export]{article}
\usepackage{longtable}
\usepackage{amssymb, amsmath, amsthm, bm, float, graphicx,  tabularx, geometry, mathrsfs, setspace, enumerate, appendix, csquotes, url, yfonts, pslatex, sectsty, eulervm, xcolor, hyperref, pdfpages, pdflscape, adjustbox}
\usepackage[T1]{fontenc}
\usepackage[round]{natbib}
\usepackage[graphicx]{realboxes}
\usepackage{pgf}
\usepackage{tikz}
\usepackage{xcolor}
\usepackage{rotating}
\usepackage{float}
\usepackage{wrapfig}
\usepackage{pdflscape}
\usepackage{longtable}
\usepackage{subfigure}
\usepackage{setspace}
\usepackage{tabulary}
\usepackage{float}
\usepackage{booktabs,caption}
\usepackage[flushleft]{threeparttable}

\begin{document}

\title{Inflation and income inequality: Does the level of income inequality matter?}



\author{Edmond Berisha\thanks{Feliciano School of Business, Montclair State University, Montclair, NJ 07043; E-mail: berishae@montclair.edu}
\and  Ram Sewak Dubey\thanks{Feliciano School of Business, Montclair State University, Montclair, NJ 07043; E-mail: dubeyr@montclair.edu}
\and Orkideh Gharehgozli\thanks{Feliciano School of Business, Montclair State University, Montclair, NJ 07043; E-mail: gharehgozlio@montclair.edu}
}
\date{\today}
\maketitle

\begin{abstract}
In the recent times of global Covid pandemic, the  Federal Reserve has raised the concerns of upsurges in prices. 
Given the complexity of interaction between inflation and inequality, we examine whether the impact of inflation on inequality differs among distinct levels of income inequality across the US states. 
Results reveal that there is a negative contemporaneous effect of inflation on the inequality which becomes stronger with higher levels of income inequality. 
However, over a one year period, we find higher inflation rate to further increase income inequality only when income inequality is initially relatively low. 

\noindent \emph{Keywords:}\; \texttt{Heterogeneity,}\; \texttt{Income inequality,}\;  \texttt{Unexpected inflation.}

\noindent \emph{Journal of Economic Literature} Classification Numbers: \texttt{D12,}\; \texttt{D30,}\; \texttt{E31.}
\end{abstract}

\setcounter{MaxMatrixCols}{10}

\newpage
\section{Introduction}

A recent resurgence of inflation in the United States has raised serious concerns about the distributional consequences of higher prices. 
Chairman of the Federal Reserve, Mr. Powell, lately noted that increases in prices are causing hardships on people living paycheck to paycheck and eating away the wealth of many investors\footnote{https://www.nytimes.com/2022/01/28/business/fed-inflation-stocks-bonds.html}. 
Motivated by these  developments in price levels and scant information about whether the impact of inflation on inequality differs across various levels of income inequality in the country, we aim, in this study, to clarify how income inequality at various quantiles responds to changes in the inflation rate. 

Earlier literature shows that the results on the impact of inflation on inequality are somewhat controversial because the relationship varies over different levels of inflation when performing cross-country analysis.
\cite{bulir1995inflation} show that inflation deteriorates income distribution in the short term. 
However, the conclusion does not hold in all cases.
For example, they find that for the United States, Finland, and Italy, rising inflation actually lowers income inequality.
For Canada, Greece, Israel, and Russia they obtain the expected "deteriorating" impact of inflation.
\cite{romer1998monetary} find that higher unanticipated inflation corresponds with higher income share for the poor and a lower Gini coefficient in the United States. 
The authors perform cross-country analysis and find that there is a negative association between poor's average income and inflation. 
They also show that the negative relationship is especially strong for industrial countries. 
\cite{fischer2000inflation} indicate that the poor are more likely than the rich to mention inflation as a top national concern.
This suggests that low income households perceive inflation as being more costly.
\cite{bulivr2001income}  show that past inflation affects current levels of income inequality. 
However, the relationship is \emph{nonlinear}. 
Decreasing inflation at a time when a country is going through an episode of hyperinflation, tends to diminish income inequality. 
Afterwards, benefits of further drop in inflation towards a very low level are found to be trivial. 
\cite{galli2001inflation} assert that the nonlinear relationship between inflation and income inequality exists because the relationship depends on the initial level of inflation. 
Specifically, the authors state that reducing inflation might decrease inequality, when inflation is initially high. 
When initial inflation is low, reducing inflation might worsen inequality. 
\cite{balcilar2018relationship} present further evidence of  nonlinear relationship between inflation and income inequality. 
They find that when inﬂation is above the threshold of 2.8 percent, it affects relative prices and increases income inequality. 

Recent studies emphasize a persistent negative relationship between the inflation rate faced by households and their income. \citet{jaravel2021} documents that the rate of inflation decreases with the household income. 
\citet{argente2020} find heterogeneity in the annual inflation rate faced by the lowest and the highest quartiles of the income distribution during the periods 2004-2007, 2008-2013 and 2014-2016. 
The difference attained its peak during the Great Recession (2008-2013) at 0.85 percentage points for 2008-2013. Surprisingly, they report very small difference at 0.02 percentage point for the post-recession period (2014-2016). \citet{hobijn2009} refine the measurement technique adopted in \citet{hobijn2005} and find a significantly negative relationship between the mean inflation and inflation inequality across households in the US. \citet{kaplan2017} consider the scanner data from the Kilts-Nielsen Consumer Panel (KNCP) for the period 2004-2013 and report that the lower income households face higher inflation with the  annual interquartile range being  6.2-9 percentage points.
\citet{gurer2020} investigate the inflation inequality in 25 European countries for a period 2001–15 and find that on average  the households in the first decile pay 11.2\% higher prices for their consumption basket compared to the households in the top decile.
This differential inflation across the income distribution contributes up to  0.04 points to the Gini coefficient in the society. 

Given the complexity of interaction between inflation and inequality, it is intuitive to postulate that the distribution consequences of inflation rates might vary as a society become more unequal over the years. 
Specifically, we contribute to the existing literature by emphasizing that the relationship between inflation and income inequality alters with higher levels of income inequality. 

Relying on the  Gini coefficient, as the measure of income inequality, we examine this hypothesis with an aim to discover how sensitive (value and sign) is the inflation coefficient, as the Gini measure ranges from the first quantile to the top, to the marginal increases in the Gini coefficient. 
We report an interesting feature of the short term dynamics of the response of Gini coefficient beginning with the current quarter to four quarters in the future. We find that the negative contemporaneous effect (first quarter) of inflation on inequality becomes stronger with higher levels of income inequality. Interestingly, we find a shift in the direction of the effect as we include further lags of inflation. Over a one year period, higher inflation rate contributes to higher inequality only when income inequality is relatively low. As level of income inequality increases, the impact of inflation become statistically insignificant.   
We focus in the U.S. because U.S. states are characterized with similar institutions and face common monetary and fiscal policies. 
This provides us with an opportunity to separate the effects of inflation on income inequality without worrying about differences on quality of institutions or policies through taxes, transfers, public sector employment, and other policy instruments.


\section{Data and Model Specification}
\subsection{Data}
We build a U.S. state-level quarterly data set that includes information on inflation, income inequality, real per capita total personal income. 
The data set is complemented by a set of fiscal and monetary policy measures, such as total state and local current government expenditures and 3-month Treasury bills. 
The inequality measure for all states is obtained from \citet{fischer2019regional}. 
The measure is constructed using household income data from the Annual Social and Economic Supplement of the Current Population Survey (CPS). 
\citet{fischer2019regional} provide extensive documentation of the construction of the inequality series. 
The authors use splines to interpolate annual measures of household income inequality to quarterly frequency. 
Inflation rate data are from \citet{hazell2020slope}. 
The measure is quarterly averages of 12-month inflation rates within a quarter year. For example, the inflation rate for 2004.Q1 is the average of 12-month inflation rates of January, February, and March of 2004, relative to the corresponding  months of 2003. 
To compute inflation rate, they use non-shelter component of the CPI\footnote{We refer the interested reader to \citet{hazell2020slope} for details on the methodology used to construct the indexes starting from micro level price data.}. The remaining variables are obtained from the Federal Reserve Bank of St. Louis database. The descriptive statistics and the source of the data employed in the analysis are summarised in Table \ref{tab:tab1} and Table \ref{tab:tab2}.

Figure 1 shows the time series of inflation rate during the sample period 1990-Q1 to 2017-Q2. Overall, there is a downward trend on the inflation rate. In the early 1990s, inflation rate was at around 4\%, followed by a decline to 2\% until late 1999. 
With the beginning of year 2000, inflation rate in U.S. began to rise again, and it reached the peak in late 2007, which is officially known as the year when U.S. economy began to slow down and enter the period of the Great Recession. 
From the beginning of the crisis, inflation followed a declining trend and stayed below 2\% until the end of the sample period with the exceptions of the  years 2011 and 2012.
During this period, inflation rate in U.S. was at around 3\%.
 
Figure 2 shows the dynamics of the Gini coefficient. Overall, there is an upward trend of income inequality in U.S. To a certain degree, we see upsurges in the Gini coefficient to correspond with lower rates of inflation.
However, there is some variation in this observed relationship. For instance, during the period 2000 to 2007, we see increases in both inflation rate and income inequality in the U.S. 
Figure 3 shows the time series of income per capita. We observe an upward trend on U.S. income per capita. 
However, whenever U.S. economy contracts, we notice drops in both the inflation and the income inequality. 
This suggest the state of the economy plays a role in driving the relationship between inflation and income inequality.
In figure 4, we plot the relationship between inflation and income inequality incorporating the size of the Gini coefficient over the years. 
On average, lower inflation corresponds with lower inequality, when inequality is initially low. 
As the level of income inequality increases, we observe the relationship to be somewhat positive. 
At the highest level of income inequality measure, the relationship between inflation and income inequality remains flat.  

We believe, the potential heterogeneous relationship between inflation and income inequality at various part of income inequality distribution can be best captured by a Panel Data Quantile Regression framework, which, in turn, permits us to provide evidence on how income inequality at different quantiles changes with higher levels of inflation. Moreover, exploring the dynamics that link inflation to inequality at a higher frequency enables us to capture and understand the social welfare costs and/or gains of inflation in short term vis-a-vis long term. 
\subsection{Model Specification}

We use a panel quantile regressions with fixed effects using the method of \cite{machado2019quantiles}. 
In the model for state $i\in\{1, 2, \cdots, I\} $ and for time $t\in\{1, 2, \cdots, T\} $ we have in generic terms:

\begin{equation}\label{eq1}
y_{it}=\alpha_{i}+X_{it}^{\prime}\beta + \left(\delta_{i} + Z_{it}^{\prime}\gamma\right)U_{it}  
\end{equation}
with $Pr\left(\delta_{i}+Z_{it}^{\prime}\gamma>0\right)=1$. 
Note that the parameters $(\alpha_{i})$ and $(\delta_{i})$ capture states fixed effects. 
$Z$ is a K-vector of known differentiable (with probability 1) transformations of the components of $X$ with element $l$ given by $Z_{l} = Z_{l}(X)$, $l = 1, \cdots, K$. 
Note that a special case of Equation (\ref{eq1}) is the linear heteroskedasticity model, in which we simply take $Z = X$, the identity transformation; a model that has been popular in the quantile regression literature (see for example, \citet{koenker1994estimatton}, \citet{he1997quantile}, \citet{zhao2000restricted}). 
Now considering 
\[
Q_{y}\left(\tau|X_{it}\right) = \left(\alpha_{i} + \delta_{i}q(\tau)\right) + X_{it}^{\prime}\beta + Z_{it}^{\prime}\gamma q(\tau),
\]
we can call $\alpha_{\tau}\equiv\alpha_{i}+\delta_{i}q(\tau)$, the $\tau$\textsuperscript{th} quantile distributional fixed effect of state $i$, which differs from the usual fixed effect in that it is not, in general, only a location shift (see \citet{galvao2011quantile}, \citet{canay2011simple}), and is allowed to have different impacts on different regions of the conditional distribution of $Y$ (refer to \citet{galvao2015efficient}, \citet{galvao2016smoothed}). 
For the details of the Method of Moments-Quantile Regression, MM-QR estimation of this model see \citet{machado2019quantiles}.

Note that in our specific application, we are interested in the relationship between inequality and inflation.
While growth of inequality, $ineq_{it}$, is our dependent variable ($y_{it}$), in the ($K=J+3$)-vector of covariates $X$ we have
\[
\sum_{j=0}^{4} \rho_{j}\pi_{it-j}+\beta_1inc_{it}+\beta_2 {ffr}_{it}+\beta_3GTE_{it}.
\]
In other words,  we consider the growth of Inequality (ineq) in this set up to depend on the lagged value, or different combinations of the lagged values of inflation ($\pi$). 
The parameter $\rho$ captures the effect of lagged values of inflation. 
Note that 
\[
\sum_{0}^{4} \rho_{j}\pi_{it-j} = \rho_{0}\pi_{it} + \rho_{1}\pi_{it-1} + \rho_{2}\pi_{it-2}+ \rho_{3}\pi_{it-3} + \rho_{4}\pi_{it-4},
\]
enables us to consider, by design, at least no lag of inflation and at most 4 lags of inflation (we can only include first lagged inflation if we set $\rho_0, \rho_2, \rho_3, \rho_4$ equal to 0, and let $\rho_1$ capture the effect of the immediate lag). 

The remaining elements of $X_{it}$ in Equation (\ref{eq1}) are the set of predictors controlling for state of the economy, monetary policy changes, and government spending. Specifically, \emph{inc} stands for income per capita; \emph{ffr} stands for federal fund rates; \emph{GTE} stands for state and local government spending.We employ growth of the control variables in our regressions.

\section{Empirical Results}

We demonstrate, graphically, the results of the estimated quantile regressions in Figures \ref{fig:fig5} and \ref{fig:fig6}.
In Figure \ref{fig:fig5}, we show the estimated coefficients of different lags of inflation on every fifth quantile of inequality measure, while controlling for income per capita. 
Note, dark blue bars (as labeled underneath the figure), provide the contemporaneous effect of inflation on every fifth quantile of inequality measure.
The green bars depict the effect of the $4$\textsuperscript{th} lag of inflation on the distribution of income inequality. 

We find the contemporaneous effect of inflation on inequality to be negative, significant, and larger as we move toward higher quantiles of income inequality.
For the lowest quantile, the estimated effect is $-0.075$, while this estimate is $- 0.177$ for the highest. 
Thus, we observe that per one unit increase in inflation, the income inequality is expected to drop by $0.08$ to $0.18$ percentage points, dependent on the level of initial inequality. 
Our findings reveal that the distributional consequences of higher inflation are sensitive to the initial level of income inequality. 

As we consider the immediate lag, the magnitude of the effect becomes smaller, still significant for some (but not all) quantiles. 
The interesting result is the shift in the direction of the effect as we include further lags of inflation. 
The third and fourth lags of inflation add positively to income inequality. 
The fourth lag of inflation, meaning the inflation of the same quarter of the previous year, shows the long term significant and positive impact of inflation on income inequality. 
One percentage point increase in inflation rate contributes to between $0.021$ to $0.136$ percent increase in inequality over $4$ quarters. 
Interestingly, the positive relationship between inflation and income inequality over a one year period is concentrated at the lower half of distribution of income inequality measure. 
Our findings imply that with higher level of income inequality, the distributional consequences of higher inflation, over a one year period, become insignificant.    

In order to have a more precise understanding of the relationship between inflation and income inequality, under different macroeconomic environments, in the regression of which the result is provided in Figure \ref{fig:fig6}, we control for differences in the monetary and fiscal policy indicators in the US, namely the federal fund rates and local and states government expenditure. 
After controlling for these indicators, we find that the long term positive effect of inflation on inequality, when inequality is initially low, is magnified.

From our results, we show evidence that inflation causes swings in income distribution rapidly. Our findings reveal that the dynamic response of inequality to changes in inflation alters over a four-quarter period. Even though the contemporaneous impact of inflation on inequality is negative, we find that over a one year period, higher inflation would exacerbate income inequality for the states that are at the bottom half of the income inequality distribution. 

\section{Sensitivity Analysis}

We follow \citet{romer1998monetary} to estimate the unanticipated inflation rate. 
Specifically, unanticipated inflation is the difference between actual inflation and forecast values from a year ago from the \emph{Survey of Professional Forecasters}. 
For example, unanticipated inflation in Q1 1990 is equal to the difference between actual inflation in Q1 1990 and one-year-ahead expectations of inflation from Q1 1989 from the \emph{Survey of Professional Forecasters}. 
Results of a similar Panel Data Quantile Regression analysis are provided in Figure \ref{fig:fig7} and \ref{fig:fig8}, in which we provide the short term and long term effect of unexpected inflation on income inequality across different quantiles of income inequality measure. 

Qualitatively, these findings are similar to the earlier results. 
We observe e negative contemporaneous affect of unanticipated inflation on income inequality. 
Interestingly, the effect remains statistically significant and similar in magnitude for the entire distribution of income inequality, except for the top quintile.     
Note that, comparing Figure \ref{fig:fig7} with Figure \ref{fig:fig5}, while the long term positive affect of unexpected inflation on income inequality prevails, the point estimates become statistically insignificant.
Our findings indicate that distributional consequences of unexpected inflation are short lived. 
These results prevail even when we control for fiscal and monetary policy changes, see Figure 8.  
 
\section{Conclusion}

Using panel quantile regression model with fixed effects, we examine  how income inequality at various quantiles responds to changes in inflation rate.
Findings suggest that the dynamic response of income inequality to changes in inflation alters over a four-quarter period and  it depends on the level of income inequality. 
Specifically, we show that the contemporaneous impact of inflation on inequality is negative and gets stronger in magnitude with higher levels of income inequality. 
However, after three quarters, the effect of inflation becomes positive and is mainly concentrated at the bottom half of the income inequality distribution. 
Interestingly, results indicate that distributional consequences of unexpected inflation are short lived.

Our findings inform policy makers that distributional consequences of higher actual and unexpected inflation occur when inequality in the society is initially low-to-moderate level.


\medskip
\bibliographystyle{plainnat}
\setlength{\bibsep}{-1pt}
\small{\bibliography{APAnonymity}}

\newpage

\begin{figure} [H]
    \centering \includegraphics[width=0.7\linewidth,height=10cm,keepaspectratio]{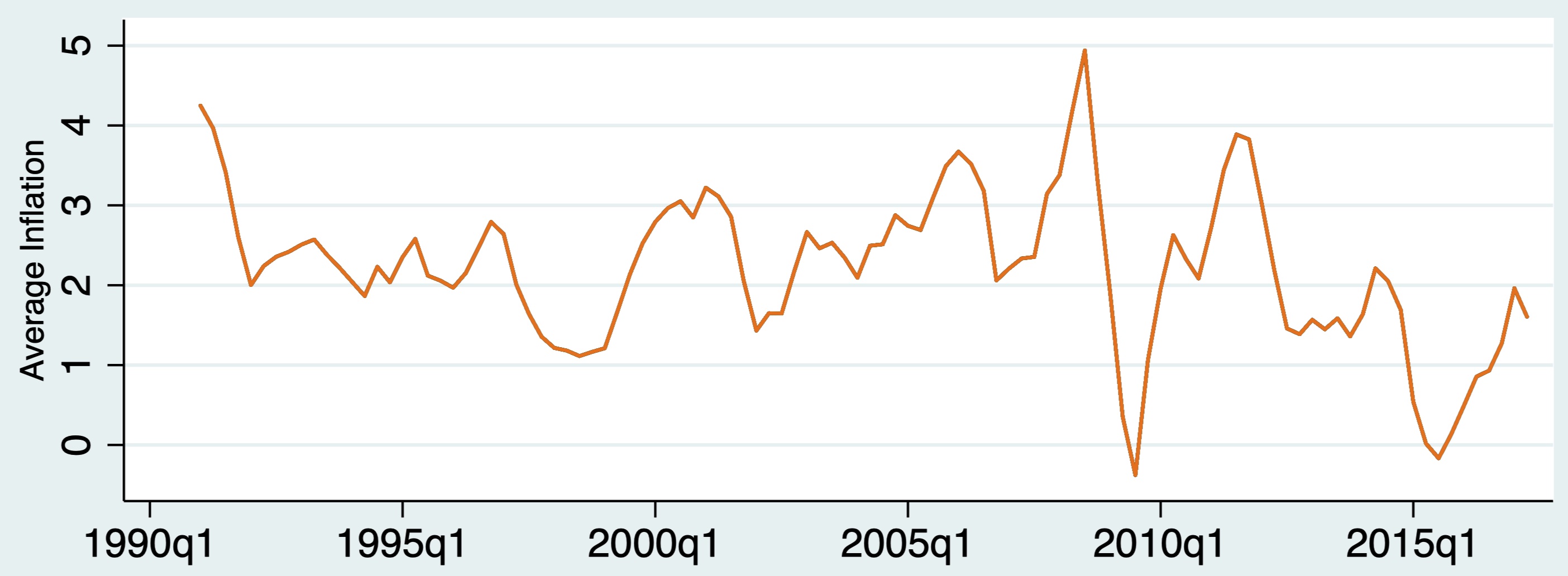} 
    \caption{\small Time Series of Inflation}
    \label{fig:fig1}
\end{figure}
\begin{figure} [H]
    \centering \includegraphics[width=0.7\linewidth,height=10cm,keepaspectratio]{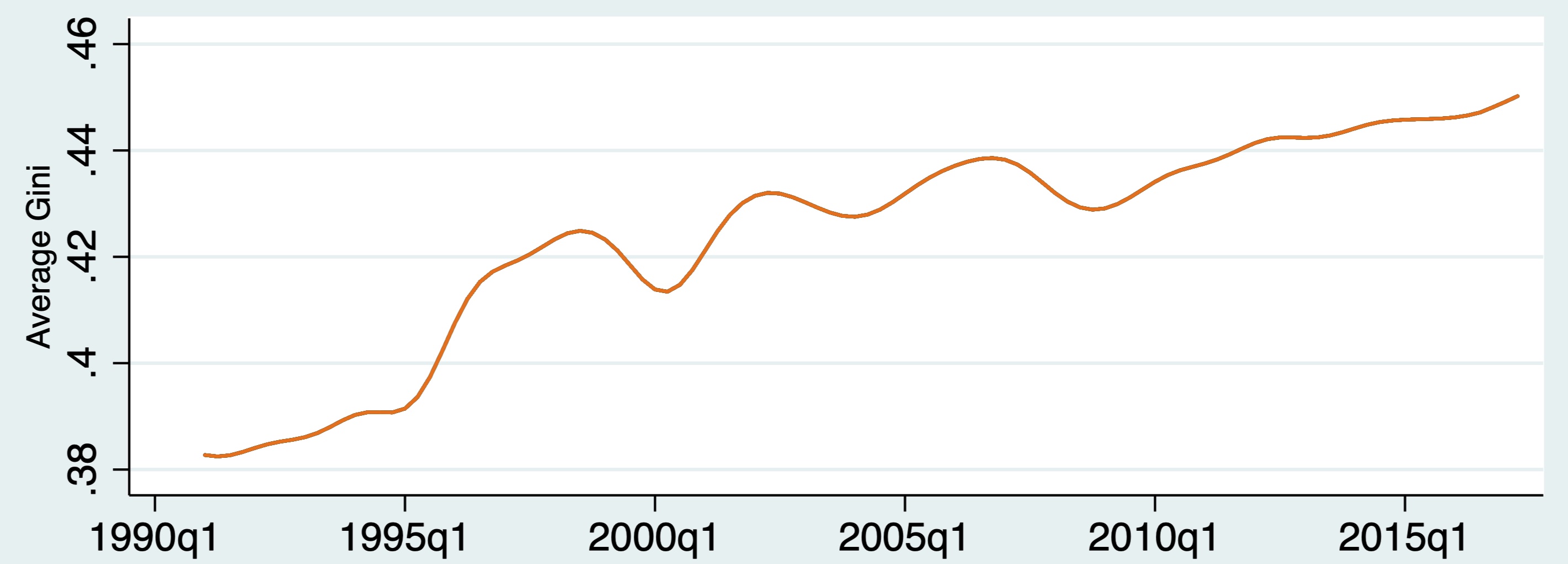} 
    \caption{\small Time Series of Gini Coefficient}
    \label{fig:fig2}
\end{figure}

\begin{figure} [H]
    \centering \includegraphics[width=0.7\linewidth,height=10cm,keepaspectratio]{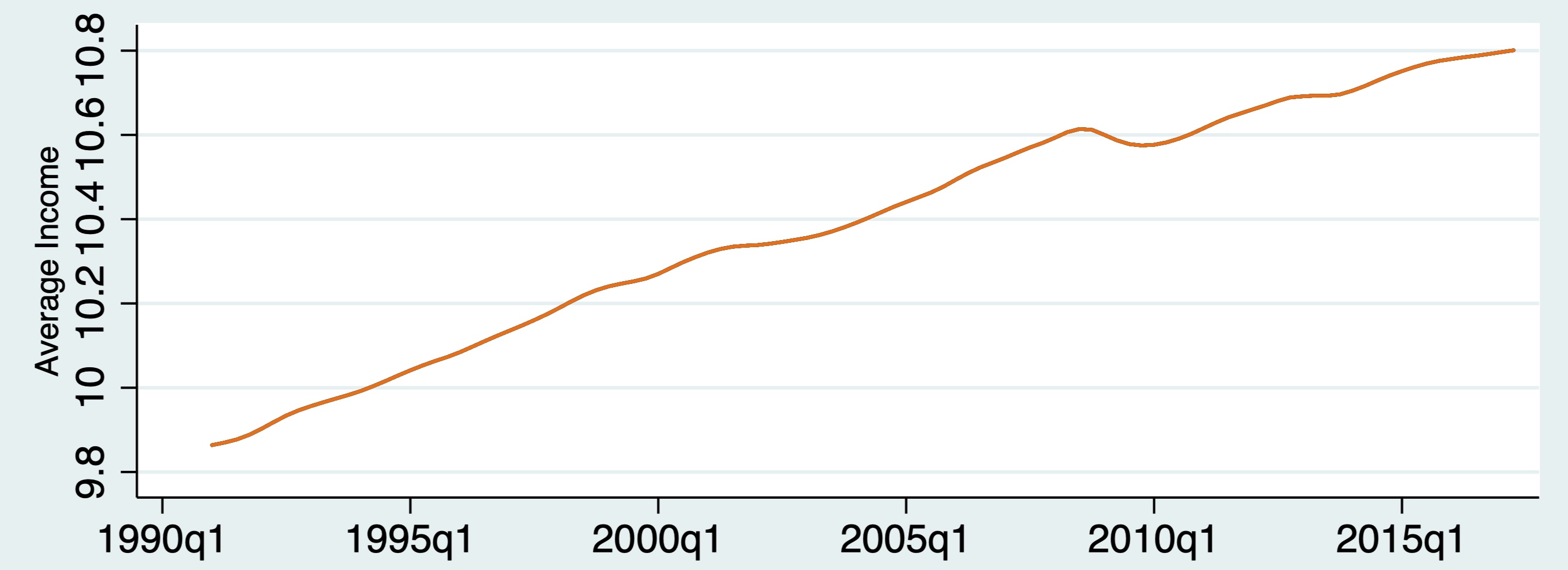} 
    \caption{\small Time Series of Income per Capita}
    \label{fig:fig3}
\end{figure}

\begin{figure} [H]
    \centering \includegraphics[width=0.7\linewidth,height=10cm,keepaspectratio]{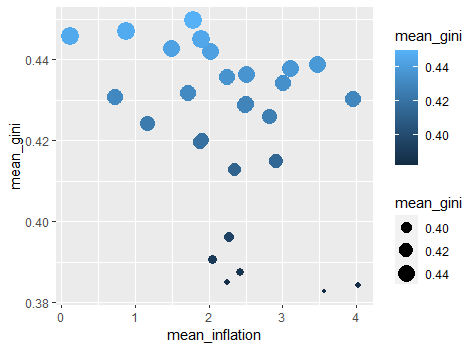} 
    \caption{\small Inflation and Inequality by State}
    \label{fig:fig4}
\end{figure}

\begin{figure} [H]
    \centering \includegraphics[width=\linewidth, height=10cm, keepaspectratio]{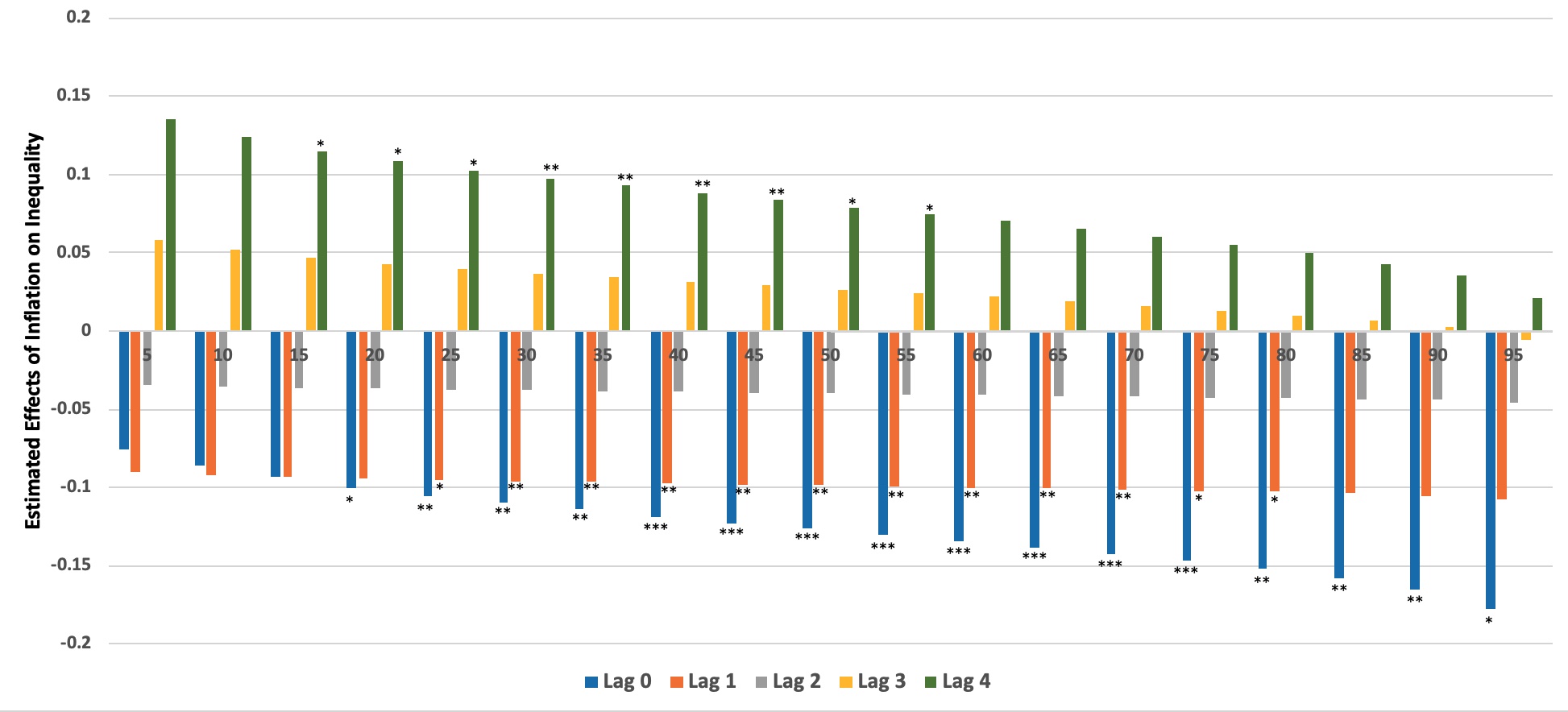} 
    \caption{\small Quantile Regression Result: 
    Estimated Effects of Inflation on Inequality, \textbf{Not Controlling} for Federal Fund Rates and Government Expenditure. 
    $\ast \left(p < 0.10\right)$, $\ast\ast \left(p<0.05\right)$, $\ast\ast\ast \left(p< 0.01\right)$}
    \label{fig:fig5}
\end{figure}

\begin{figure} [H]
    \centering \includegraphics[width=\linewidth,height=10cm,keepaspectratio]{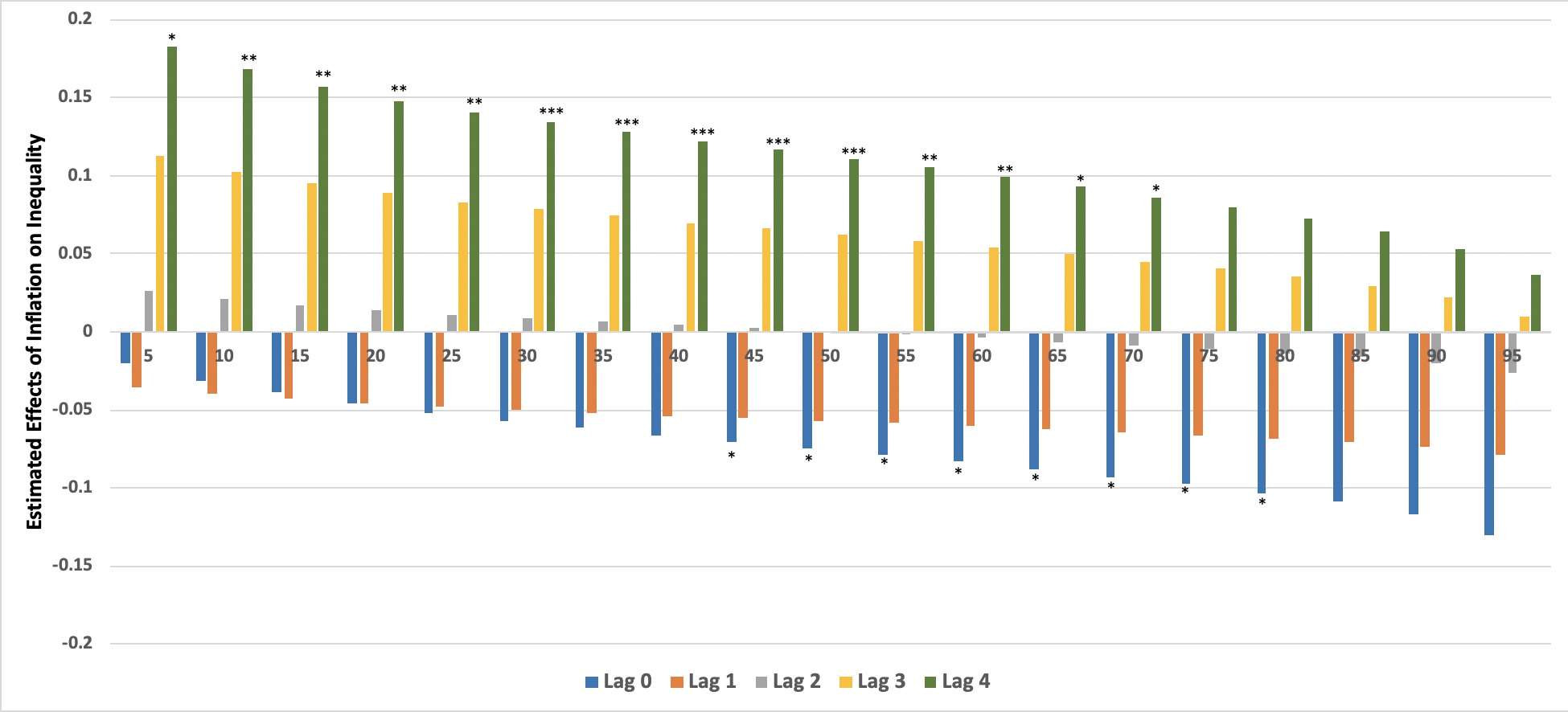} 
    \caption{\small Quantile Regression Result: Estimated Effects of Inflation on Inequality, \textbf{Controlling} for Federal Fund Rates and Government Expenditure. 
    $\ast \left(p < 0.10\right)$, $\ast\ast \left(p<0.05\right)$, $\ast\ast\ast \left(p< 0.01\right)$}
    \label{fig:fig6}
\end{figure}

\begin{figure} [H]
    \includegraphics[width=\linewidth,height=10cm,keepaspectratio]{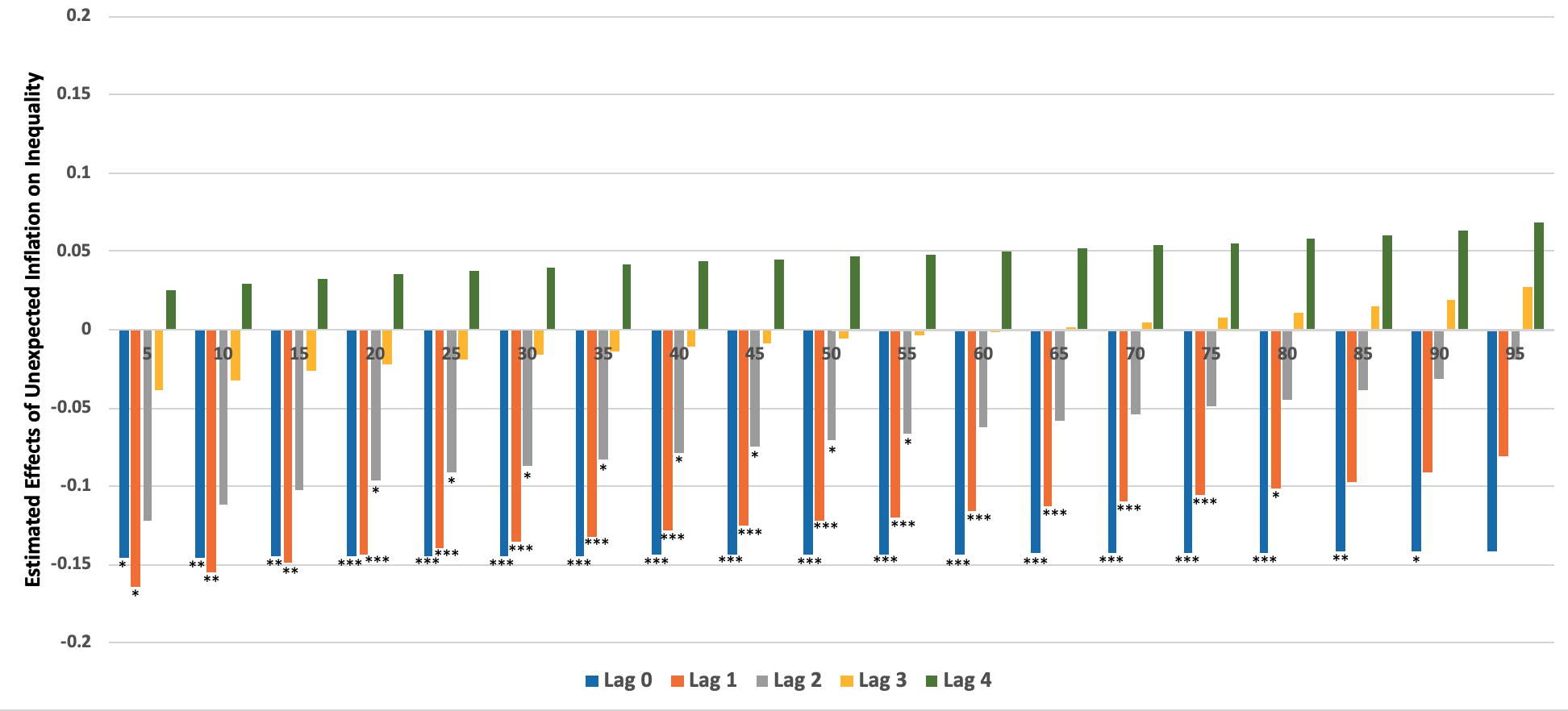} 
    \caption{\small Quantile Regression Result: Estimated Effects of \textbf{Unexpected} Inflation on Inequality, \textbf{Not Controlling} for Federal Fund Rates and Government Expenditure. 
    $\ast \left(p < 0.10\right)$, $\ast\ast \left(p<0.05\right)$, $\ast\ast\ast \left(p< 0.01\right)$}
    \label{fig:fig7}
\end{figure}

\begin{figure} [H]
    \centering \includegraphics[width=\linewidth,height=10cm,keepaspectratio]{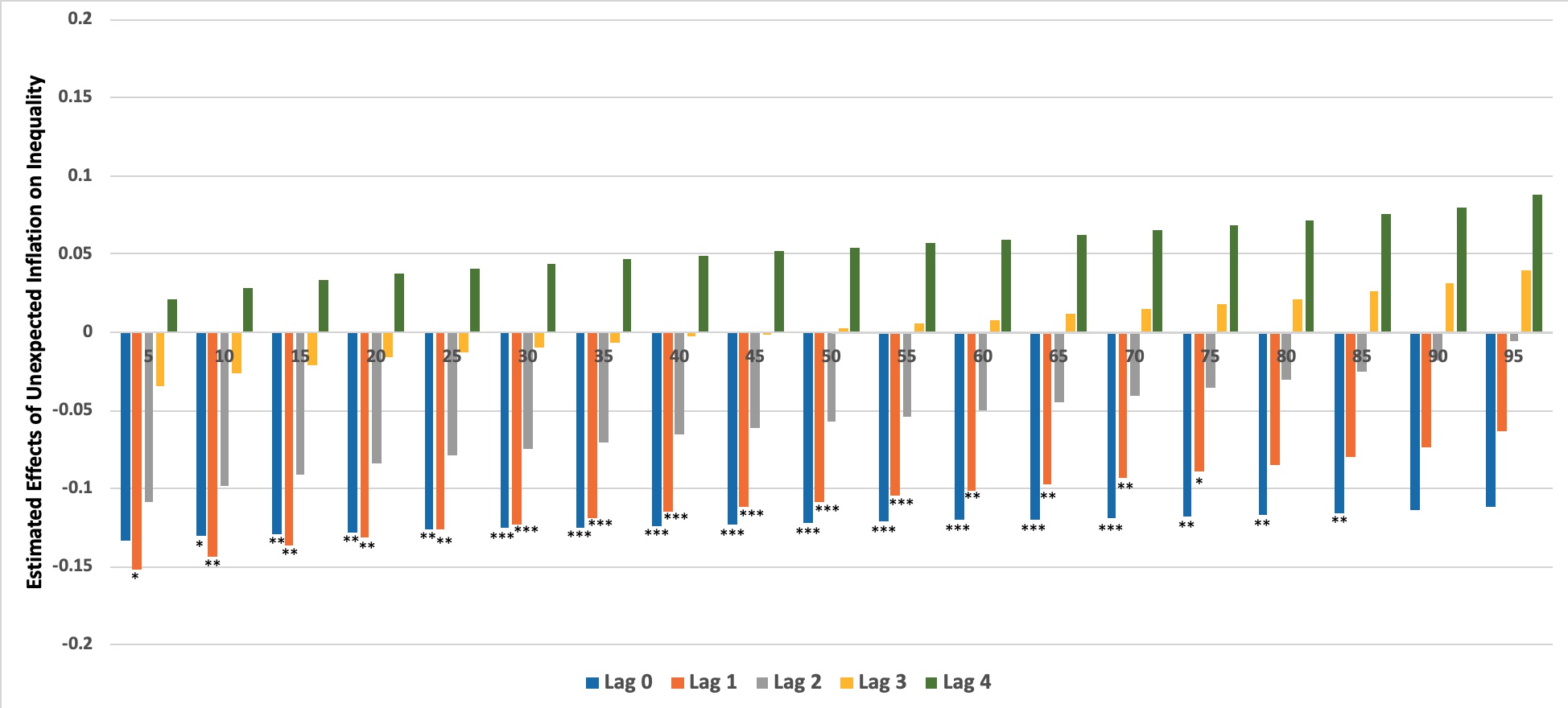} 
    \caption{\small Quantile Regression Result: Estimated Effects of \textbf{Unexpected} Inflation on Inequality, \textbf{Controlling} for Federal Fund Rates and Government Expenditure. 
    $\ast \left(p < 0.10\right)$, $\ast\ast \left(p<0.05\right)$, $\ast\ast\ast \left(p< 0.01\right)$}
    \label{fig:fig8}
\end{figure}

\newgeometry{margin=2cm}
\begin{sidewaystable}
\caption{Summary of the data sources employed for the analysis} \label{tab:tab1} 
\def\arraystretch{1}
\ignorespaces 
\centering 
\footnotesize
\begin{tabular}{m{5em} |m{15em}| m{5em}| m{20em}}\hline\hline
Variable&   Definition  &  Abbreviation used &   Source\\\hline
Inequality Growth  &  The measure is constructed using household income data from the Annual Social and Economic Supplement of the Current Population Survey (CPS). Change from a year ago  &  inecgrowth &    \cite{fischer2019regional}  \\\hline
Inflation &  The measure is quarterly averages of 12-month inflation rates within a quarter year. For example, the inflation rate for 2004.Q1 is the average of 12-month inflation rates of January, February, and March of 2004, relative to the corresponding months of 2003.    &    pi &  \cite{hazell2020slope}\\\hline
Income Growth   &  Income per capita,  Change from year ago.  & incgrowth   &   \cite{fischer2019regional}    \\\hline
Federal Fund Rate  &   Federal Fund Rate,  Change from year ago, percent.  & dffr & Federal Funds Effective Rate, retrieved from \href{https://fred.stlouisfed.org/series/DFF}{FRED-DFF} \\\hline
Government Expenditure  & Total Local and State Government Expenditure,  Change from year ago& dGTE & U.S. Bureau of Economic Analysis, State and Local Government Current Expenditures, retrieved from \href{https://fred.stlouisfed.org/series/SLEXPND}{FRED-SLEXPND}\\\hline
Unexpected Inflation  & Unanticipated inflation
is the difference between actual inflation and forecast values from a year ago.    &  unexpingfl&     \cite{romer1998monetary}\\\hline\hline
\end{tabular}
\end{sidewaystable}
\restoregeometry

\newpage

Table \ref{tab:tab2} provides descriptive statistics. 
\begin{table*}[!htbp]
 \caption{\textbf{Descriptive Statistics}}
  \label{tab:tab2} 
\def\arraystretch{1}
\ignorespaces 
\centering 

\begin{tabular}{l|c|c|c|c|c|c}
\hline\hline
            &            &        Mean&   Std. Dev.&         Min&         Max&   N/n/T-bar\\
\hline
ineqgrowth  &  overall   &    .56&    3.14&    -11.07&     12.35&        3740\\
            &  between   &           .&    .19&    .10&    1.00&          34\\
            &   within   &           .&    3.13&   -11.26&    12.32&         110\\
            \hline
pi          &  overall   &    2.28&    1.38&   -4.03&    9.35&        3740\\
            &  between   &           .&    .23&    1.82&    2.90&          34\\
            &   within   &           .&    1.36&   -3.72&    9.21&         110\\\hline
incgrowth   &  overall   &    3.63&    2.90&   -105.82&    12.08&        3740\\
            &  between   &           .&    .27&    2.77&    4.20&          34\\
            &   within   &           .&    2.89&   -104.95&    11.63&         110\\\hline
dffr        &  overall   &   -.31&    1.40&   -4.34&    2.59&        3740\\
            &  between   &           .&           0&   -.31&   -.31&          34\\
            &   within   &           .&    1.40&   -4.34&    2.59&         110\\\hline
dGTE        &  overall   &    4.60&    2.63&    -1.38&    11.98&        3740\\
            &  between   &           .&           0&    4.60&    4.60&          34\\
            &   within   &           .&    2.63&    -1.38&    11.98&         110\\\hline
unexpinfl   &  overall   &   -.34&     1.42&   -6.27&     5.49&        3740\\
            &  between   &           .&    .23&   -.80&    .28&          34\\
            &   within   &           .&    1.40&   -5.83&    5.45&         110\\\hline
union       &  overall   &    .47&    .50&           0&           1&        3740\\
            &  between   &           .&    .51&           0&           1&          34\\
            &   within   &           .&           0&    .47&    .47&         110\\
\hline\hline
\end{tabular}

\par 
\end{table*}

\end{document}